\documentclass[useAMS,usenatbib]{mn2e}
\usepackage{graphicx}
\usepackage{amsmath}

\title[P--P scattering and
the formation of debris disks]
{Impact of planet--planet scattering on   
the formation and survival of debris disks}
\author[F. Marzari]{F. Marzari$^{1}$ \\
$^{1}$Dept. of Physics, University of Padova, 35131 Italy}

\begin{document}

\date{Accepted .....;  Received ..... ; in original form ........}

\pagerange{\pageref{firstpage}--\pageref{lastpage}} \pubyear{.....}

\maketitle

\label{firstpage}

\begin{abstract}
Planet--planet scattering is a major dynamical mechanism 
able to significantly alter the architecture of a planetary system. In addition 
to that, it may
also affect the formation and retention of 
a debris disk by the system. A violent chaotic evolution of the planets
can easily clear 
leftover planetesimal belts preventing the ignition of a substantial
collisional cascade that can give origin to a debris disk.
On the other end, a mild evolution with limited steps in 
eccentricity and semimajor axis can trigger the formation of 
a debris disk by stirring an initially quiet planetesimal belt. 
The variety of possible effects that planet--planet scattering
can have on the 
formation of debris disks is 
analysed and the statistical probability of the different 
outcomes is evaluated. This leads to the prediction that 
systems which underwent an episode of chaotic 
evolution might have a lower probability of harboring
a debris disk. 

\end{abstract}


\section{Introduction} 
\label{intro}

Most of the belts of dust or debris that have been detected 
around many main sequence stars 
like HR4796A \citep{jura91}, AU Mic \citep{kalas04},
HD107146 \citep{williams04} as well as our sun,
are thought to originate from a ring
of planetesimals, leftover of the planet formation process. 
If a population of large solid bodies have formed within the 
protoplanetary disk, as predicted by the core--accretion model,
at later stages of the system evolution, when the star has 
reached
its mature state, they may collide 
creating smaller and smaller "debris" dust grains. The 
collisional cascade of these remnant planetesimals
effectively replenish the dust population long after it would have 
normally dispersed because of P-R drag, further 
collisions or interaction with planets (see \cite{wyatt08} for a 
review). 
The 
constant refilling of dust due to cratering and fragmentation 
within the planetesimal belt maintains at present 
what we observe as a debris disk surrounding the star.
To produce dust by collisions, the disk must be stirred enough 
to ignite the collisional cascade. Different mechanisms have been 
proposed to activate a planetesimal belt and these include 
stellar flybys, self stirring and planet stirring (see 
\cite{matt14} for a review). On the other hand, the planetesimal 
belt must survive in order to continuously replenish the dust disk and, 
as a consequence, a smooth and limited evolution of the planets
after their formation is required. 
Planet migration caused by tidal interaction 
with the gas of the native protoplanetary disk \citep{kley07} or 
planetesimal--driven migration \citep{murray98,levison07} 
both may lead to changes 
in the planetesimal orbital distribution but, in most 
cases, they would not totally clear remnant planetesimal belts from the 
system. Even a mildly violent evolution, such as that suggested for 
our solar system by the Nice model \citep{tsiganis05}  and characterized by 
a significant outward migration of Neptune and Uranus driven by 
close encounters with Saturn, possible sculpted the Kuiper belt
but did not cleared it out. 

Different is the scenario in those extrasolar systems harboring 
planets in highly eccentric orbits. 
The leading mechanism for explaining why extrasolar planets
have very elliptic trajectories is planet--planet
scattering \citep{rasio-ford96, weiden-marza96,
MW02,juritre08,naga08,
CHATTERJEE08, rayarmgo09, rayarmgo2010, marzari10,
naga11, rayall2011, rayall2012,
benesv12}, hereinafter P-P scattering.  Often called the ‘‘Jumping Jupiters''
model, it assumes that two or more massive giant
planets form from the disk around a solar--type star.
Mutual gravitational perturbations among the planets 
excite their eccentricities leading to
dynamical instability and crossing orbits. Repeated close 
encounters between the planets cause an extended period of chaotic 
evolution characterized by major changes in the orbital
configuration. The most likely
outcome is the ejection of one (or more) planet on a hyperbolic trajectory,
leaving one (or more) planets on a stable eccentric and inclined 
orbit. 
The orbit of the inner survivor is closer to
the star than the innermost starting orbit, since it supplies 
orbital energy to the ejected planet(s). 

During the chaotic evolution preceding the ejection of one planet 
on a hyperbolic trajectory, the planets roam around the system 
on eccentric and inclined orbits for an extended period. An 
important question concerning the final aspect of these 
systems is: can 
planetesimal belts formed prior to the onset of instability 
survive this violent phase? In other words, is the probability of
detecting debris disks the same in planetary sytems which 
underwent P--P scattering compared to those in which 
this kind of  dynamical evolution did not occur? If indeed 
P--P scattering tends to clear planetesimal belts then
we would expect 
a lower rate of debris disk 
detections around stars with planets on 
eccentric orbits. On the other hand, if the chaotic 
evolution of planets does not clear potential planetesimal 
belts, it would leave them in a dynamically excited state sufficient to 
trigger the collisional cascade needed to form the dust disk.
In this case, all systems which underwent P--P scattering
would have stirred planetesimal disks and the debris disk
statistics would be richer.  

While \cite{bonsor2013} investigated the short--lived production 
of exozodiacal dust just after a P--P scattering event 
due to an increase of mass scattered in inner orbits from 
outer Kuiper belts and \cite{rayarm2013} studied the formation 
of mini--Oort clouds, 
here the impact of the chaotic phase on the long term 
survivial of debris disks is explored. Initially 
unstable planetary systems 
populated by three planets 
on circular orbits and by a non--excited swarm 
of leftover planetesimals 
is numerically integrated in time during the P--P
scattering evolution until it ends and the two
surviving planets achieve stable orbits.
By inspecting the final orbital distribution of the 
remaining planetesimals 
it is possible to predict whether they are able
to generate and maintain a debris disk or, instead, whether
they are too scattered 
to produce a dusty disk dense enough to be observed. 
Once identified the main types of final dynamical states of 
the two stable planets and of the surviving planetesimal rings
the next important step will consist in 
drawing some statistical predictions about the presence of 
debris disks in aged planetary systems which, in their 
early evolution,  underwent a
P--P scattering event. 

\section{``Jumping Jupiters'' and planetesimals belts: the initial 
set up}

To explore the impact of P--P scattering on 
the formation and survival of debris disks, 
this study focuses on planetary systems initially 
populated by
three Jupiter--size planets on circular orbits 
and by a leftover planetesimal belt.
The inner planet has a fixed semimajor axis of 3 AU while the 
outer two bodies are placed on orbits with
semimajor axes close 
to the stability limit
\citep{MW02,CHATTERJEE08,marzari10,marza14}.
The values of $a_2$ and 
$a_3$ are computed as $a_2= a_1 + K_1 \cdot R_{H_{1,2}}$ and $a_3 = a_2 +
K_2 \cdot R_{H_{2,3}}$ where $R_H$ is the mutual Hill's sphere,
$K_1$ is a random number ranging from 2.8 to 3.4 
while $K_2$ is encompassed between  4 and 5.8. 
The choice of this range for $K_1$ and $K_2$ is dictated by the need of 
simulating systems that become unstable on a short timescale
(lower than 10 Myr) for computational reasons. Equal 
values for $K_1$ and $K_2$,
as shown in \cite{marza14},  are not really
required for exploring the P--P scattering occurrence in 3--planet 
systems. 
The initial 
eccentricities of the planets are randomly selected between 
0 and 0.01 while the inclinations range from $0^o$ to 
$0.5^o$. The remaining orbital angles are randomly drawn from
$0^o$ to $360^o$. 64 different systems made of 3 planets and a planetesimal 
belt have been integrated.

Two different planetesimal belts are separately 
modeled, an inner belt ranging from 1 to 30 AU  and an 
outer one ranging from 30 to 60 AU (similar to the Kuiper Belt). 
We assume that these belts are not initially excited and have
small eccentricities and inclinations 
($e < 0.001$ and $i<0.5^o$). 2500 massless bodies are integrated 
for each belt and their semimajor axis is selected randomly 
between the belt limits. This produces a radial density distribution
that declines as $r^{-1}$ but the final dynamical outcome can 
be easily rescaled to any initial radial distribution. The 
RA15 version of the RADAU numerical 
integrator \citep{ever85} has been used to compute the orbital evolution of 
the bodies. The choice has been dictated by its stability and precision 
when dealing with gravitational encounters of and with Jupiter--size bodies. 

50 simulations have been 
performed with 3 equal--mass planets where $m=1 M_J$ while in 
14 we considered planets with different masses $m=1 M_J$, $m=2 M_J$
and $m=3 M_J$.

\section{Chaotic evolution of the planets: effects on the planetesimal 
remnant population}
 
The survival of an asteroid belt or of a Kuiper belt in a planetary 
system where P--P scattering occurred depends on 1) how long and wild 
is the evolution of the planets prior the ejection of one of them 
and 2) on the final architecture 
of the system i.e. where the outer planet
is placed at the end of the chaotic behavior. Both these aspects 
may be very different and, since the evolution is fully chaotic, 
cannot be predicted a priori.
In the next sections different final 
outcomes will be discussed 
ranging from dispersal of the belt, partial excitation 
and almost no effect. This last case is typical of those 
systems where P--P 
scattering leads to the ejection of one planet on a very short 
timescale.  A statistical prediction of the impact of the 
different outcomes on the presence of debris disks will be 
given in the next section on the basis of the outcome of a large number of
P--P scattering test events. 

\subsection{Planetesimal dispersal: no debris disks}

\begin{figure}
\resizebox{90mm}{!}{\includegraphics[angle=0]{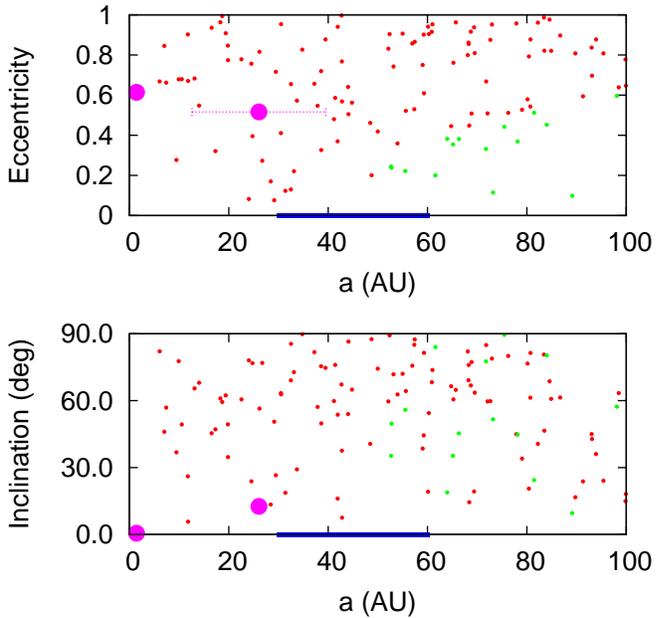}}
\vskip -4.1 truecm
\caption{Final orbital distribution of an initially 
quiet planetesimal belt 
after an extended period of P--P scattering. The 
blue dots are the initial 2500 planetesimals
located in between 30 and 60 AU and similar to a Kuiper Belt. 
The initial semimajor axis of the two outer planets are 
$a_2=3.88$ AU and $a_3=6.10$ AU.
After 2.1 Myr of chaotic evolution, the ring is significantly 
dispersed. The large magenta filled circles are the surviving planets 
and the error bar marks the limits of the radial excursion of the bodies 
due to their eccentricity. The red dots are those planetesimals 
crossing the planet orbits at the end of the P--P scattering event
while the green ones are non--interacting planetesimals that may survive
for a long period if they are not trapped in chaotic regions related to 
resonances. 
}
\label{f1}
\end{figure}

In Fig.\ref{f1} the final orbital distribution of an initial 
quiet planetesimal belt extending from 30 to 60 AU is shown 
after the end of the 
chaotic evolution of the planets. One body is ejected on a 
hyperbolic trajectory and the two surviving planets are left on 
stable eccentric orbits both inside the initial location of the
belt. The P--P scattering phase has lasted 2.1 Myr
and in this period about 95\% of the planetesimals are 
ejected out of the system. Those which are still orbiting the 
star are in a 
highly excited state with eccentricities reaching
almost 1, inclinations up to $90^o$ and a few also on 
retrograde orbits. Most of them, due to  
their high eccentricity, are doomed to encounter the outer planet,
which is also on a higly eccentric orbit, 
and be ejected out of the system later on. 
Only a few planetesimals located beyond 
50 AU do not interact with the planets and may survive for a 
long time. However, their number density is much  
lower than the initial one and it appears unlikely that
they can activate a significant collisional cascade able to 
create and maintain a debris disk also because of their large
dispersion in eccentricity and inclination. This farther 
decreases the volume density preventing an  
intense collisional activity. 

\begin{figure}
\resizebox{90mm}{!}{\includegraphics[angle=0]{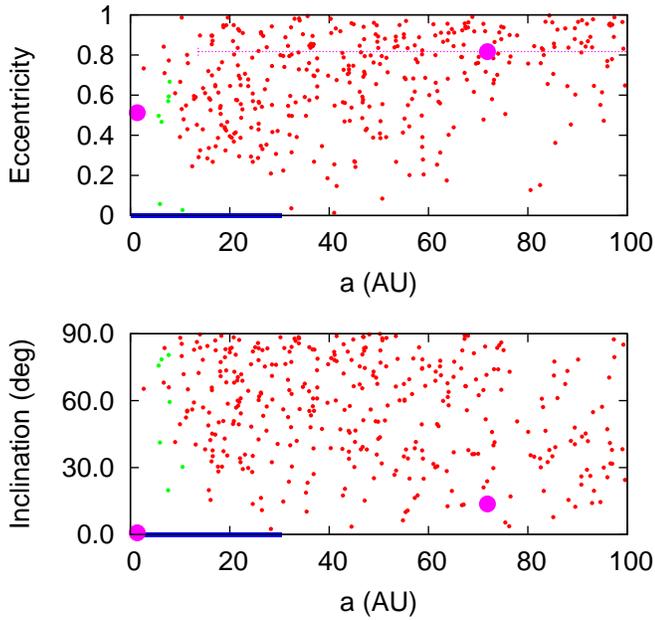}}
\vskip -4.1 truecm
\caption{Same as  Fig.\ref{f1} but $a_2=3.92$ AU and 
$a_3= 6.12$ AU. The P--P scattering phase is 
shorter and the outer planet is inserted in a highly eccentric 
orbit. Its gravitational perturbations will fully clear any 
planetesimal belt located within 130 AU 
preventing the formation of a 
significant debris disk. 
}
\label{f2}
\end{figure}

A second event which may prevent the formation of a debris disk 
after a P--P scattering phase is the 
insertion of the outer planet into a highly eccentric orbit. 
This outcome is illustrated in Fig.\ref{f2} where the outer planet,
whose semimajor axis is moved to about 72 AU at the end of 
the chaotic phase, 
has an eccentricity of $\sim 0.8$ that allows it to cover 
a wide range in radial distance.  
The initial planetesimal belt, located in this case
between 1 to 30 AU, is stirred up during the P-P scattering 
phase which lasts about 0.9 Myr and when the planets are injected 
in their final orbits it is destined to be fully cleared. 
Even this kind of dynamical outcome is adverse to the 
formation of a debris disk in the system. This second kind of 
P--P scattering evolution is lethal also for a potential 
Kuiper Belt since the planet reach extends well beyond 
100 AU clearing any leftover planetesimal belt
initially present in the system. 

\subsection{Planetesimal stirring: activation of debris disks}

\begin{figure}
\resizebox{90mm}{!}{\includegraphics[angle=0]{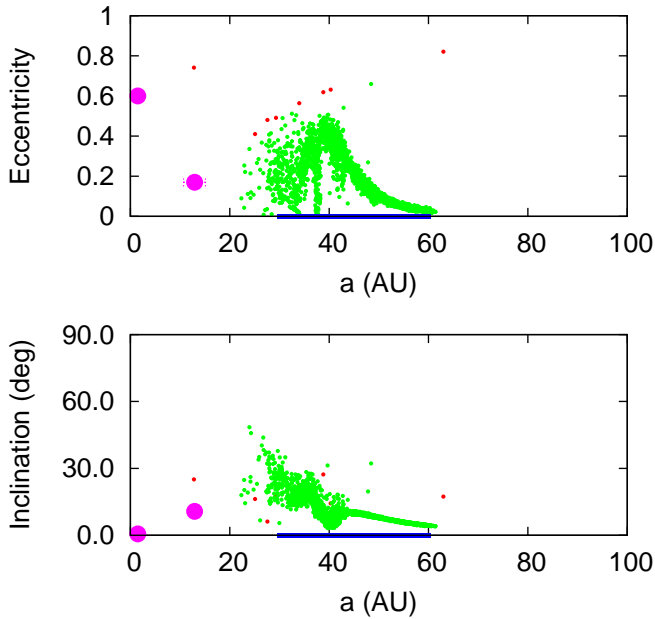}}
\vskip -4.1 truecm
\caption{Activation of a debris disk by stirring a 
belt of planetesimals in the outer regions of the system. 
The P--P scattering in this case does not clear the 
planetesimal ring but it excites higher values of 
eccentricity and inclination triggering a high 
dust production rate.  In this case 
$a_2 = 3.93$ AU and $a_3 = 6.23$ AU.
}
\label{f3}
\end{figure}

\begin{figure}
\resizebox{90mm}{!}{\includegraphics[angle=0]{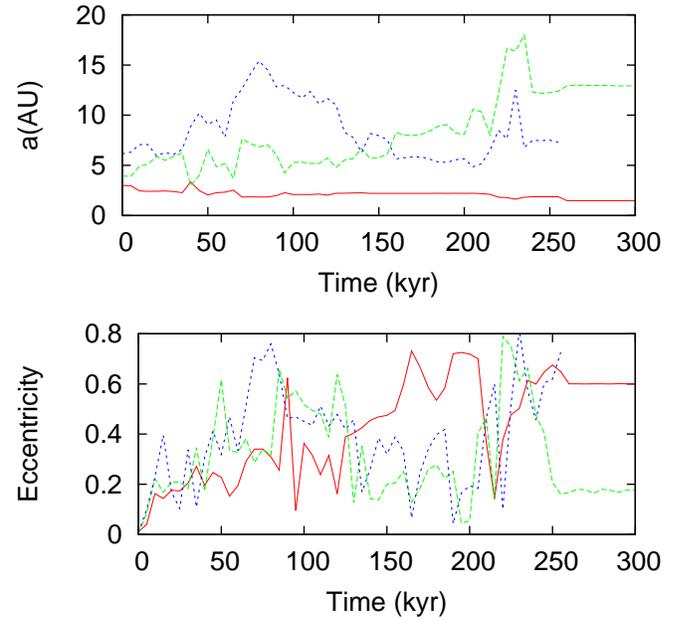}}
\vskip -4.1 truecm
\caption{Evolution of the planets during the P--P scattering phase 
that has lead to the excitation of the belt illustrated in Fig.\ref{f3}.
The semimajor axes of the planets never jump beyond 20 AU and 
the outer planet achieves a low final eccentricity. 
}
\label{f3bis}
\end{figure}

\begin{figure}
\resizebox{90mm}{!}{\includegraphics[angle=0]{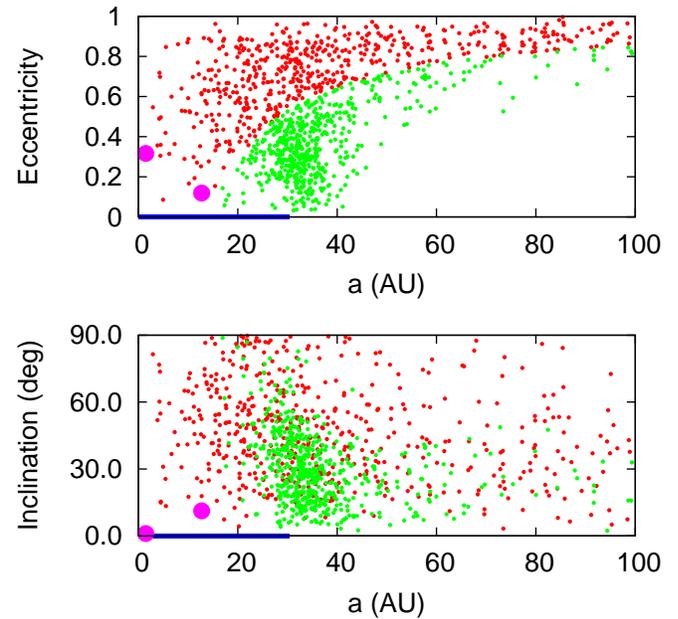}}
\vskip -4.1 truecm
\caption{An excited planetesimal belt
is created at the end of the P--P scattering phase
($a_2 = 3.90$ AU and $a_3=5.93$ AU). The 
green dots, representing planetesimals that do not cross
the planet orbits, extend beyond 20 AU. 
}
\label{f4}
\end{figure}

The cases described in the previous section are characterized either
by an extended period of chaotic evolution or by the insertion of 
the outer planet in a highly eccentric orbit. However, the P--P scattering 
may occur on a short timescale and the final eccentricity of the 
surviving planets may be low. In this case, the chaotic evolution of 
the planets is an important source of velocity stirring able to 
ignite the collisional cascade leading to a debris disk without clearing
the belt. This behavior is illustrated in Fig.\ref{f3} where an
initial Kuiper Belt is efficiently stirred up during the P--P scattering
phase. The chaotic evolution of the planets, prior the ejection of one 
of them, is limited in semimajor axis and it is shown in Fig.\ref{f3bis}.
The two surviving planets are left on low eccentricity 
orbits allowing a debris disk to form and be refilled. A somewhat 
more excited belt is produced in the case illustrated in Fig.\ref{f4}. 
The initial planetesimal ring is stirred up by encounters with 
the planets and finally a stirred belt is formed extending from about 
20 AU and beyond. High velocity collisions are dominant and dust 
is produced.

The type of evolution described in this section 
is characterized by 
a mild chaotic evolution of the planets with contained changes 
in eccentricity and semimajor axis. It is highly favorable to 
the development of a debris disk from a planetesimal belt, better
if located in the outer regions of the system. It excites the 
planetesimal trajectories leading to high velocity impacts 
with a consistent rate of dust production able to refill 
the debris disk. Different degrees
of stirring are possible and they depend on the 
timespan of the planet chaotic evolution and on the 
final orbital elements of the planets. 

\subsection{No excitation} 

In some cases, the P--P scattering phase is very short and 
a planet is ejected almost immediately on a timescale of some 
thousands years. In this case the amount of stirring of a 
leftover planetesimal belt is negligible unless one of the
planets directly perturbs the belt. The planetary
system would be able to excite the planetesimals only via secular 
perturbations or resonances. 
\vskip 0.6 truecm 

It is noteworthy that the 14 cases we considered with 
planets having different masses ($m=1 M_J$, $m=2 M_J$
and $m=3 M_J$) show a dynamical behaviour morphologically very similar 
to those with equal mass planets. Thereby, they are not 
commented in this Section but they will be considered from 
a statistical point of view in the next Section.

\section{Statistics of the P--P scattering and its implication 
for the formation of debris disks.}

In the previous section the different degrees of 
perturbation of a P--P scattering period on planetesimal 
belts have been discussed. In some cases the belt is 
dispersed while in others it is only stirred up leading 
to the activation of a debris disk. Due to the chaotic nature 
of the evolution of planets in crossing orbits, it is not
possible to predict a priori the effects on 
planetesimal belts of the P--P scattering phase. However, 
it is possible to have a rough glimpse at the frequency with which 
the different cases can occur.  To achieve this goal, the initial 
configuration with 3 planets on unstable orbits has been 
replicated  with the same criteria adopted to generate the initial 
orbital elements described in Sect. 2. 
Due to the strong chaotic nature of the planet evolution
after the first close encounter,  
the choice of a limited range of values for $K_1$ and $K_2$
but random values for all other orbital elements allow to span 
the whole spectrum of possible final dynamical states of the system
once fixed $a_1$ and the masses of the planets. To cover a 
wider scenario of possible systems, we have performed three
additional statistical simulations where 
also different masses and wider initial orbits have 
been considered. These 
models are motivated by the presence of a significant number 
of exoplanets with masses beyond that of Jupiter. In spite of 
a decreasing trend 
beyond 1 $M_J$,
there is a consistent number of bodies with masses
of 2 and 3 $M_J$ on eccentric orbits which may have
been involved in a P--P scattering event.
It is also 
impossible, at present, to predict the orbital distribution 
of a multi--planet system before the onset of a
P--P scattering phase since it depends on the initial 
density of the protoplanetary disk, the amount of orbital 
migration of the planets during their growth, 
if the P--P scattering event occurs in presence or 
not of the disk gas and on many additional factors. For these reason, 
a different intial 
architecture for the 3--planet systems is studied with the whole 
system moved outwards in semimajor axis setting $a_1 = 5$ AU. 
In these statistical simulations, each system is
numerically propagated until one planet is ejected. The orbital elements
of the surviving planets are recorded in order to test the 
frequency with which the different outcomes occur. 2000 
3--planet systems are integrated in each different case.

\begin{figure*}
\begin{minipage}[c]{.40\textwidth}
\hskip -1.0 truecm
\resizebox{90mm}{!}{\includegraphics[angle=-90]{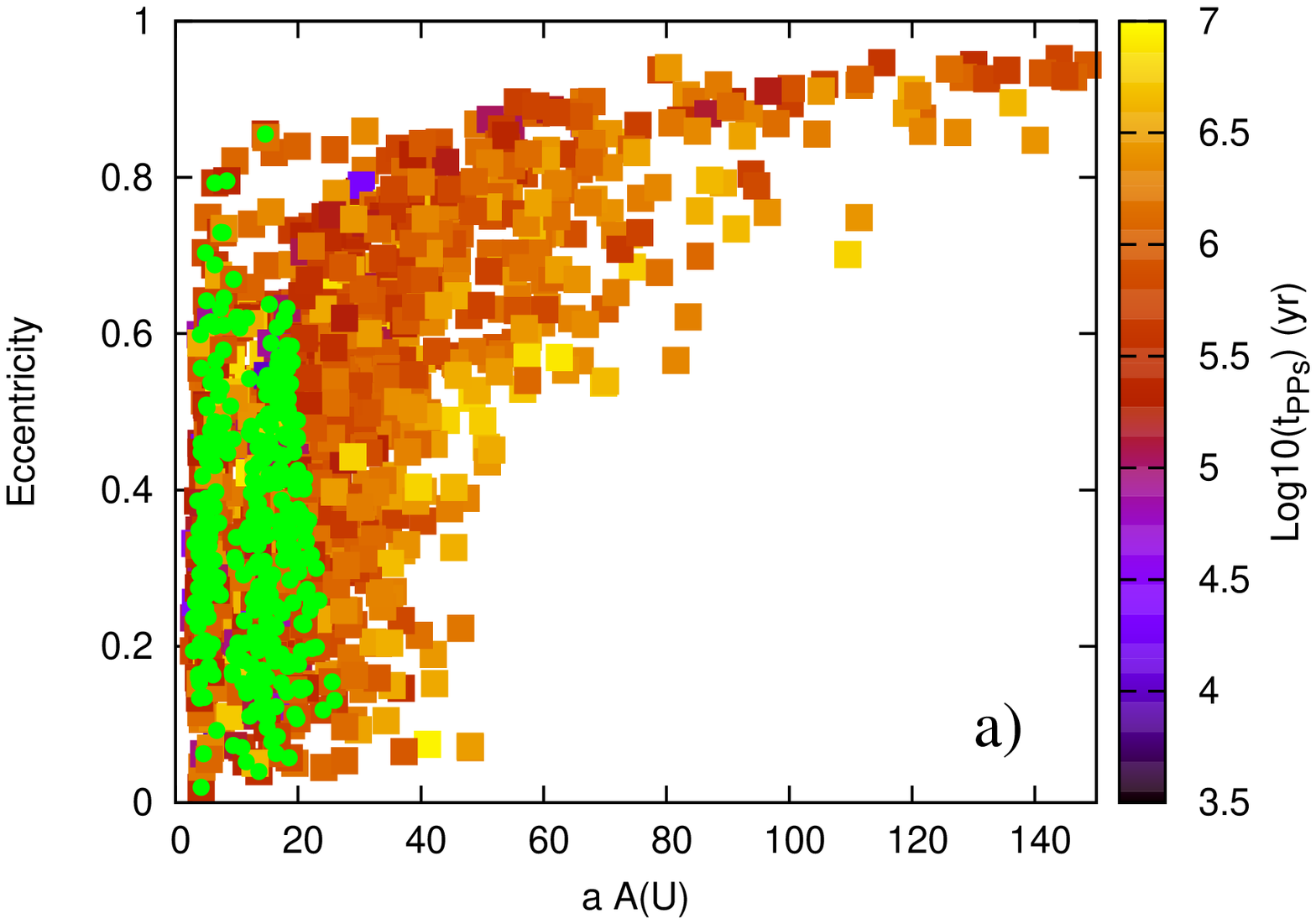}}
\end{minipage}%
\hspace{10mm}%
\begin{minipage}[c]{.40\textwidth}
\resizebox{90mm}{!}{\includegraphics[angle=-90]{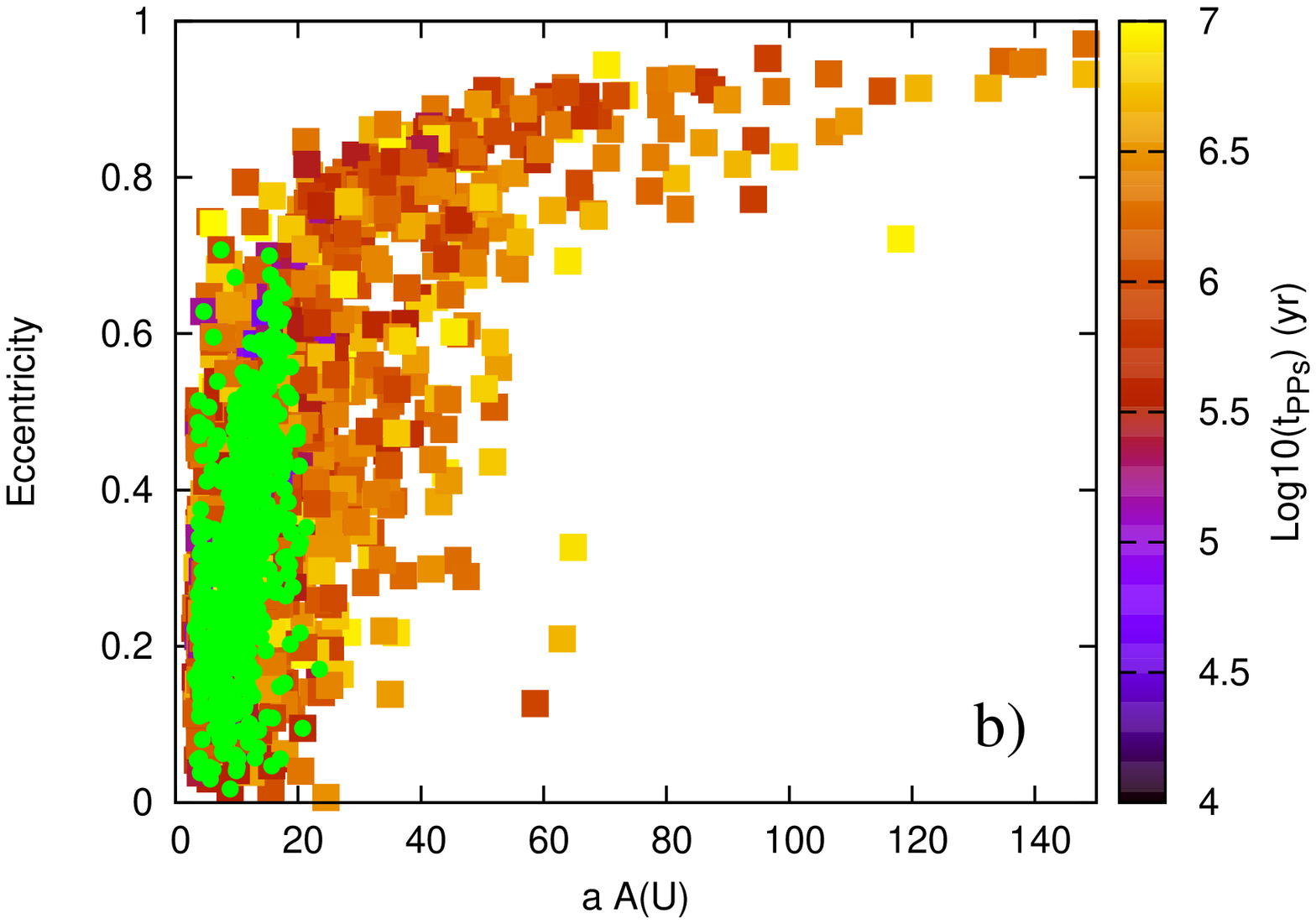}}
\end{minipage}%
\hfill
\begin{minipage}[c]{.40\textwidth}
\hskip -1.0 truecm
\resizebox{90mm}{!}{\includegraphics[angle=-90]{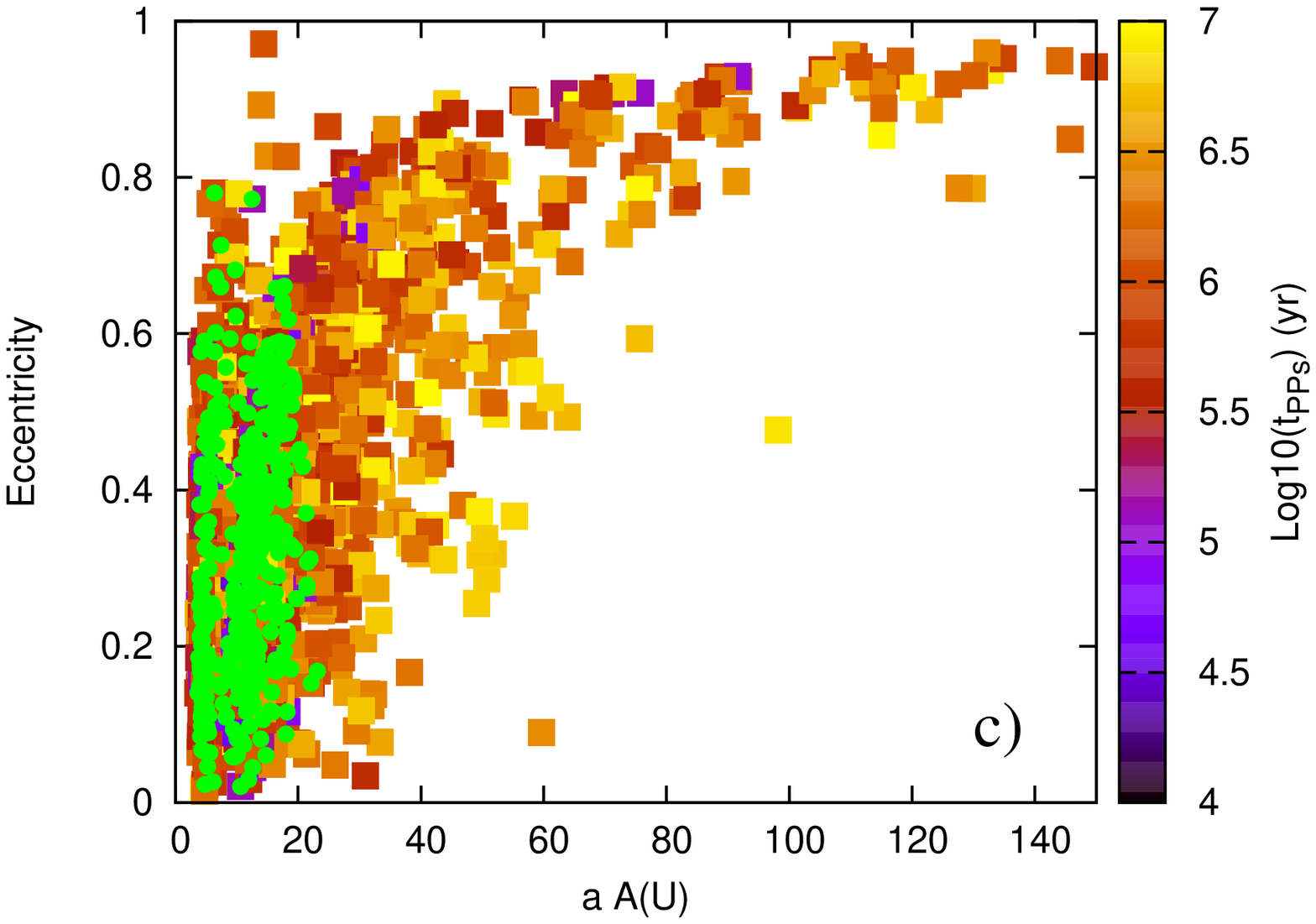}}
\end{minipage}%
\hspace{10mm}%
\begin{minipage}[c]{.40\textwidth}
\resizebox{90mm}{!}{\includegraphics[angle=-90]{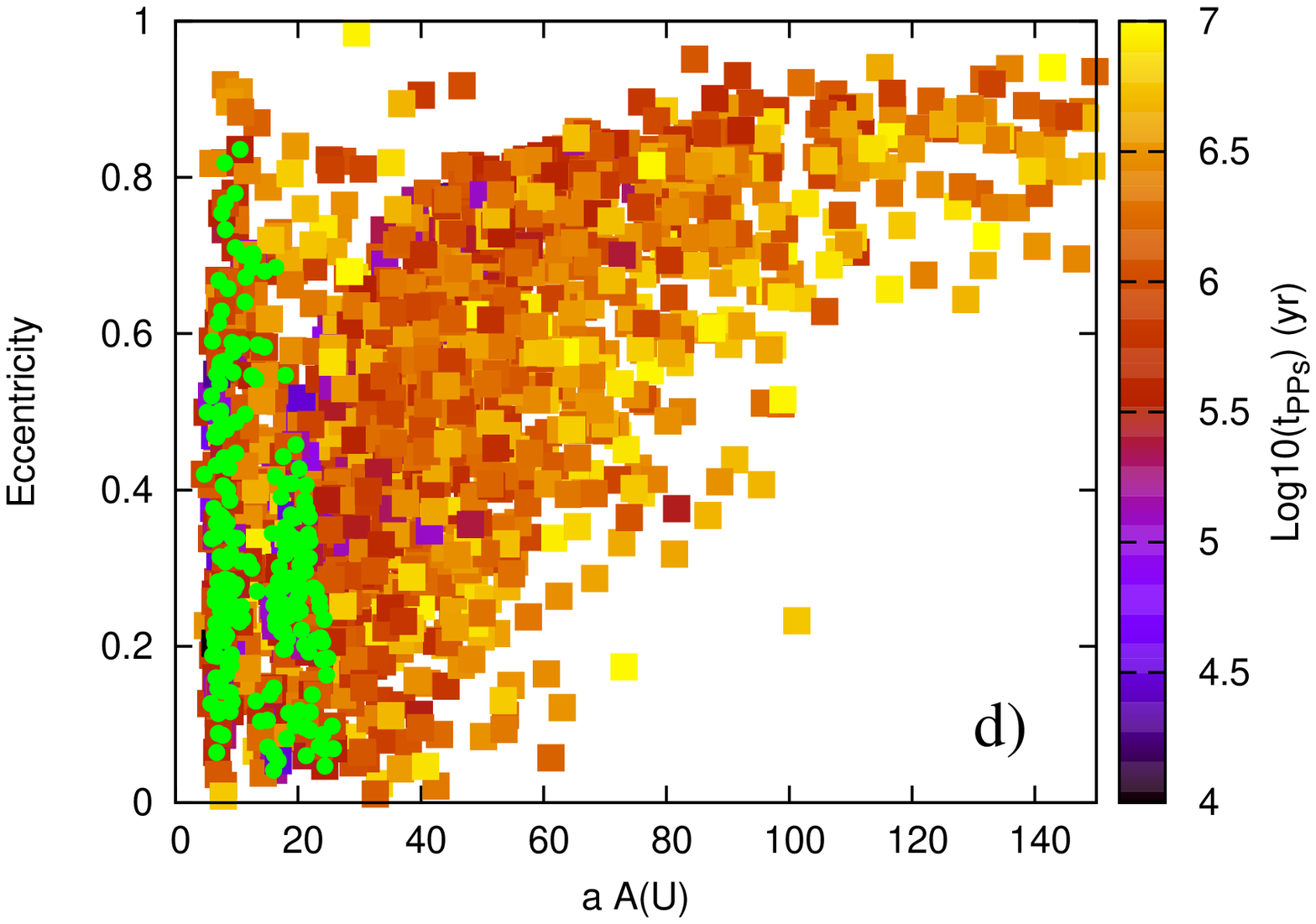}}
\end{minipage}
\caption{Statistical distribution of the outer planet
orbital elements at the end of the P--P scattering period. 
The color coding shows the logarithm of the timespan 
of the chaotic phase, from the onset of instability to the 
ejection of one planet. The green filled circles mark 
those systems where the timespan of the chaotic evolution is
shorter than $1 \times 10^6$ yr and the aphelion of the 
outer planet is lower than 30 AU. These configurations 
should be favorable to the ignition of a debris disk. 
In panel a) the masses of the planets are $m_1 = m_2 = m_3 =
M_J$, in panel b) $m_1 = 1M_J$ ,  $m_2 = 2 M_J$ and   $m_3 =
3 M_J$, in panel c) $m_1 = 2 M_J$,  $m_2 = 1 M_J$ and  $m_3 =
2 M_J$ while in panel d) the masses of the 3 bodies are all
equal to $1 M_J$ while the semimajor axes are 
$a_1=5 AU$, $a_2= a_1 + K_1 \cdot R_{H_{1,2}}$ and $a_3 = a_2 +
K_2 \cdot R_{H_{2,3}}$, 
respectively.
}
\label{f5}
\end{figure*}

In Fig.\ref{f5}a
we show the final distribution of the outer planet semimajor
axis vs. eccentricity for the case with $m_1=m_2=m_3 = 1 M_J$
(case a). 
The color coding gives the timespan of the 
P--P scattering phase which is important in determining the 
amount of dispersal of the planetesimal belt. 
In this plot we can draw a rough separation 
between highly perturbing systems, that possibly inhibit the formation 
of a dusty disk, and those where, on the contrary,  the mild perturbations 
favor the formation of a disk. The green dots mark those sytems 
where the timespan of the chaotic evolution is shorter than 
$1 \times 10^6$ and the aphelion is lower than 30 AU. The majority 
of these systems should activate potential leftover Kuiper--like 
belts and have debris disks, at least populating the outer region 
of the system. All the other systems are potentially hostile to 
the survival of any planetesimal belt and should not posses 
debris disks. It is noteworthy that the large majority of systems
(76\%) belong to this second category and are expected 
not to harbor a debris disk since its potential precursors 
have been eroded away by the perturbations 
of the planets either during an extended chaotic phase or 
because the outer planet is on a highly eccentric orbit. 
In Fig.\ref{f5}b the masses of the planets are set to
$m_1= 1 M_J$, $m_2= 2 M_J$, and $m_3 = 3 M_J$, respectively. For this 
initial configuration (case b) we observe an increase of systems 
which may 
harbor debris disks since the fraction of planets on outer and
eccentric
orbits is lower compared to the standard case a) with 3 equal mass
planets.  In model b)  62\% of the planetary sytems are hostile 
environments for the formation of a debirs disk, a lower 
percentage compared to 
case a). When the planet masses are set to 
$m_1= 2 M_J$, $m_2= 1 M_J$, and  $m_3 = 3 M_J$
(Fig.\ref{f5}c) the situation is somewhat intermediate 
between case a) and b) with the number of systems potentially 
adverse to debris disk development up to 71\%. The most critical 
scenario is that shown in case d)
where the planets, all with the same mass equal to $M_J$,
are shifted on wider orbits ($a_1=5 AU$).
It is noteworthy in this case the proliferation of systems with the outer 
planet on a wide and eccentric orbit and, as a consequence, 
87\% of the systems 
appear to be hostile to the formation of a debris disk. 

\section{Discussion and conclusions}

A large fraction of extrasolar planetary systems harbor planets
on highly eccentric orbits and relatively small semimajor axes. 
The most promising mechanism to 
explain this finding is that the planetary system had an 
additional planet which was ejected after a period of dynamical
instability. The P--P scattering phase that precedes the ejection 
of one planet on a hyperbolic trajectory naturally leads to the onset
of highly eccentric and inclined orbits for the 
surviving planets explaining observations. On the long term, 
the inner planet, if close to the star, would have its orbit
circularized by stellar tides. The outer one would preserve 
its high eccentricity. 

An important question concerning these systems is the fate of 
remnant planetesimal belts, leftover of the planet formation 
process, and their capability of forming and maintaining 
debris disks in spite of the P--P scattering event. 
Two conditions are strongly unfavorable to the survival of 
these belts: an extended period of chaotic behavior 
before the ejection of one planet and the insertion of 
the outer surviving planet in a highly eccentric orbit.
In the first case the prolonged chaotic evolution of the 
planets characterized by large steps in semimajor axis and eccentricity 
bring them frequently within the belt where they excite and 
scatter a large fraction of the planetesimals that populate it. 
When finally the P--P scattering comes to an end, most of the 
belt is cleared and the chance of forming a debris disk is 
very low due to the large orbital dispersion of the surviving 
planetesimals. In the second case, when the outer planet is ejected 
in a high eccentricity orbit, it spans a wide range of radial 
distance scattering out of the system every body it encounters 
on its path. It disperses even potential Kuiper Belts if the 
eccentricity is high enough. 

If the chaotic evolution evolve on a short timescale and the outer
planet ends up on a orbit which is not too eccentric, then 
planetesimal belts can survive the P--P scattering event 
and are left in an excited state which easily activates the
collisional cascade leading to a 
debris disk. The disk can either be related to an asteroid belt or, more 
frequently, to a Kuiper Belt. 

In a minority of cases, the P--P scattering will occur on such a short
timescale that the planets wll not have the time to excite the 
belt during the chaotic evolution and, unless the outer planet 
is in a very eccentric orbit, the belts in the system will not 
be stirred up. In this case, other dynamical mechanisms are
required to trigger high velocity collisions, similar to those 
in systems where P--P scattering did not occur \citep{matt14}.

A statistical exploration of the P--P scattering event in terms 
of timespan of the chaotic evolution and orbital elements of 
the outer surviving planet shows that  the fraction of  
events leading to a highly perturbing configuration depends 
on the dynamical architecture of the 3 planets 
prior the onset of the chaotic phase. A different mass distribution 
among the planets or their shift to wider orbits may either enhance
or reduce the number of cases potentially hostile to the 
survival of a planetesimal belt. In particular, if the planets 
enter the chaotic phase when they are far from the host star, 
like 
in the case d) described in Section 4, then the erosion of 
remnant planetesimal belts is expected to occur in the vast majority 
of cases. 
A general conclusion,
that can be drawn from the statistical analysis performed 
in Section 4, 
is 
that from about 60 to 90\% of systems which underwent a 
period of P--P scattering potentially cleared their 
leftover planetesimal belts.  
This suggests
that planetary systems where at least one planet has been 
discovered on a highly eccentric orbit, probable outcome of 
a P--P scattering period, should have a lower rate of debris 
disks. 

The analysis presented here needs to be extended to 
other different 
initial configurations for the planets. Systems with only 2 
initial planets or with more than three should be investigated
to give more stringent predictions about the fate of 
planetesimal belts and debris disks. 

\section*{Acknowledgments}
We thank an anonymous referee for his useful comments and suggestions. 

\bibliographystyle{mn2e}
\bibliography{paper}

\bsp

\label{lastpage}

\end{document}